# Sculpting harmonic comb states in terahertz quantum cascade lasers by controlled engineering

Elisa Riccardi,[1] M. Alejandro Justo Guerrero,[1] Valentino Pistore,[1] Lukas Seitner,[2] Christian Jirauschek,[2] Lianhe Li,[3] A. Giles Davies,[3] Edmund H. Linfield,[3] and Miriam S. Vitiello[1*]

*[1] NEST, CNR - Istituto Nanoscienze and Scuola Normale Superiore, Piazza San Silvestro 12, 56127, Pisa, Italy*
*[2] TUM School of Computation, Information and Technology, Technical University of Munich (TUM), Hans-Piloty-Str. 1, 85748 Garching, Germany*
*[3] School of Electronic and Electrical Engineering, University of Leeds, Leeds LS2 9JT, UK*

**Optical frequency combs (FCs), that establish a rigid phase-coherent link between the microwave and optical domains of the electromagnetic spectrum, are emerging as a key high-precision tools for the development of quantum technology platforms. These include potential applications for communication, computation, information, sensing and metrology, and can extend from the near-infrared with micro-resonator combs, up to the technologically attractive terahertz (THz) frequency range, with powerful and miniaturized quantum cascade laser (QCL) FCs. The recently discovered ability of the QCLs to produce a harmonic frequency comb (HFC) – a FC with large intermodal spacings – has attracted new interest in these devices for both applications and fundamental physics, particularly for the generation of THz tones of high spectral purity for high data rate wireless communication networks, for radiofrequency arbitrary waveform synthesis, and for the development of quantum key distributions. The controlled generation of harmonic states of a specific order remains, however, elusive in THz QCLs. Here we devise a strategy to obtain broadband HFC emission of a pre-defined order in QCL, by design. By patterning *n* regularly spaced defects on the top-surface of a double-metal Fabry-Perot QCL, we demonstrate harmonic comb emission with modes spaced by (*n+1*) free spectral range and with a record optical power/mode of ~270 μW.**

Quantum cascade lasers (QCLs)[1] emitting at terahertz frequencies (1.2–6 THz)[2,3] are unipolar semiconductor lasers characterized by ultrafast gain recovery dynamics (~20 ps)[4], and a large third-order $\chi^{(3)}$ Kerr nonlinearity.[5,6] These features, combined with their small footprint, intrinsic spectral purity,[7] continuous wave (CW) operation and relatively high output powers (> 2 W)[8], make these sources an ideal platform for the observation of mode-locked multimode regimes[9] and quantum correlation effects.[7] The experimental discovery that QCLs can spontaneously emit optical frequency combs (OFCs)[9–11] has dramatically increased their technological potential in metrology,[12] communication,[13] sensing,[14] dual-comb imaging and spectroscopy,[15,16] and quantum technologies.[14,17,18]

OFCs operating in the harmonic regime (harmonic frequency combs (HFCs)) represent a promising technology.[19,20] In contrast to the standard evenly spaced comb operation where the mode separation matches the cavity round trip frequency, in the harmonic state, optical power is concentrated in a few modes spaced by several multiples of the cavity free spectral range (FSR).[21] For both QCL FCs and HFCs, quantum correlations are expected to be established by resonant four-wave-mixing (FWM)[9,14,22,23] enabled by the third-order $\chi^{(3)}$ Kerr nonlinearity, which is the mechanism responsible for QCL comb mode proliferation and for frequency- and phase-locking the growing side modes.[5,6] FWM has been also identified as the core mechanism of entanglement among laser modes in passive micro-resonator-based comb emitters,[24] triggering an increased interest for the exploitation of the same effect in standard Fabry-Perot (FP) QCL FCs.

Under pure electrical injection, the frequency spacing of THz QCL combs is fixed by the length of the cavity and by the effective group index of the waveguide.[25] The two types of coherent regime, FC and HFC, can, however, co-exist in the same device for different values of the *dc* current, as has been observed in both mid-infrared (mid-IR)[19,26] and THz[27] QCLs[19,28]. In general, coherent regimes alternate with irregular unlocked states upon sweeping the bias current.[29] However, experimental access to HFC states in electrically driven THz QCLs has been limited to date. The only mechanism to control these devices is through the injected current. Usually, the QCL first reaches a state of high single-mode intracavity intensity, and when this intensity is large enough, an instability threshold is reached caused by the $\chi^{(3)}$ population pulsation nonlinearity, favouring the appearance of modes separated by tens of free spectral ranges (FSR)s from the first lasing mode.[27]

Controlled generation of harmonic states of a specific order has been only achieved in mid-IR QCLs either by the injection of an optical seed into the laser cavity,[30] or, through defect engineering in mid-IR ring QCLs[31]. However, similar attempts remain elusive in THz QCLs.

Here, we propose an experimental methodology that alleviates the influence of device fabrication and operation uncertainties on the emission state of the FC, allowing the specific order of the HFC to be determined in advance. We engineer, by design, broadband HFCs with a tailored number of harmonic modes of pre-defined order in a standard FP cavity THz QCLs. By employing a heterogeneous gain medium[32,33], and a cavity architecture that integrates a fixed number $n$ of scattering points (defects) placed at designed points along the top-surface of the double-metal QCL cavity, we demonstrate harmonic comb emission over the entire operational range of the QCL, with harmonic modes spaced by $(n+1)$ FSR. Our approach adopts a standard linear cavity, rather than a ring cavity, and does not require external modulation which would necessitate complex external circuitry. The ability to control HFC generation by design in THz QCLs paves the way for novel applications, such as the generation of multi-mode squeezed states of light, arising from quantum-correlated side-band modes.

Fig. 1a shows the schematic diagram of a prototypical device, comprising an edge-emitting double-metal waveguide QCL, of length $L$, with $n = 2$ defects patterned on the top contact in the form of narrow rectangular slits covered by a thin layer (5 nm) of nickel.
The equidistant defects divide the cavity into $n+1$ parts spaced by $L/(n+1)$, creating a periodic pattern on the top contact. This affects the spatial dependence of the intermode beat intensity along the cavity, originating from the interference of any combination of lasing frequency comb modes.

The optical modes sustained by the resonator are determined by $L$, as in any FP structure, but the sequence of patterned defects acts as a frequency filter that introduces periodic additional losses. This leads to an intensity modulation of the frequency of the modes supported by the cavity. We perform three-dimensional simulations via a finite element method (Comsol Multiphysics) to evaluate the effect induced by the surface patterning on the $TM_{00}$ QCL modes. Details of the simulation parameters are provided in the Supplementary Information. The simulations show that the losses induced by the defects strongly depend on defects' spatial position with respect to the field phase. Fig. 1b displays the electric field distribution inside the QCL cavity with one defect placed in the centre of the cavity ($L/2$). This design is intended for the generation of a second harmonic FC. The optical modes peak at the metallic cladding on the top of the active region. Two prototypical neighbouring modes of the comb with frequencies of 3.009 THz and 2.992 THz are shown in the top and bottom panels of Fig. 1b, respectively. They correspond to a harmonic and a non-harmonic mode, respectively. The black rectangle and dashed line between the panels indicates the defect position on the top contact of the cavity. In the first case (top), the defect position is aligned with a peak of the electric field, while in the second case (bottom), the defect and the field peaks are misaligned. We can evaluate the effect induced by the defects on the electric field by looking at the redistribution of the field's components. Due to the intersubband nature of the device, the photons propagating in the cavity are transverse magnetic (TM) polarized, i.e. with the electric field parallel to the growth direction ($z$). The defects redistribute the electric field components, thereby reducing the $E_z$ field amplitude and introducing additional losses to the modes whose electric field maximum/minimum is not aligned with the defect (see Supplementary Information).

To quantify the field redistribution introduced by the defects, we can extract the ratio $E_z/E_{tot}$, where $E_{tot} = E_x+E_y+E_z$ is integrated over the entire cavity. The comparison between the reference structure without defects shows that $E_z/E_{tot}$ drops, on average, by 0.7% when the defects are all aligned with the field peaks, while it drops by 11% for the frequencies for which the defects and field peaks are misaligned, i.e. the non-harmonic modes. This field redistribution resembles a mode-dependent radiative loss modulation. We calculate the radiative losses through the 3D quality factor ($Q$ factor), extrapolated as a direct output of the simulations (see Supplementary Information). If we express the harmonic modulation as the ratio of the average radiative losses between the harmonic and non-harmonic modes, we find that by introducing a defect in the centre of the cavity, the radiative losses are almost the same for the harmonic modes, but they are 20% higher for the non-harmonic ones. Finally, upon introducing a lossy material, e.g. a thin nickel layer covering the defect, the losses increase for all modes, but particularly for the nonharmonic ones, leading to an increase of the harmonic modulation of 22.3% (see Supplementary Information).

We next simulate the spectral response of the defects on the cavity modes in a range of 20 FSR around 3 THz, i.e. the central frequency of the QCL employed here.[32,33] Figs. 1c – 1f show the normalized power spectral density (PSD) calculated in the presence of an increasing number of defects. The corresponding geometry is illustrated in the insets at the top of each panel. Fig. 1c shows the results obtained in a standard FP cavity without defects, and acts here as our reference. In Figs. 1d-f, we show the simulated results for a waveguide divided by equidistant defects, with $n$ = 1 (Fig. 1d), 2 (Fig. 1e), and 5 (Fig. 1f). In the defect free FP cavity (reference), the simulated PSD reveals a series of equidistant modes of comparable intensity, as expected. Conversely, when the defects are implemented, the PSD presents a visible frequency-dependent modulation that affects all comb modes except those at the $n$-th harmonic.

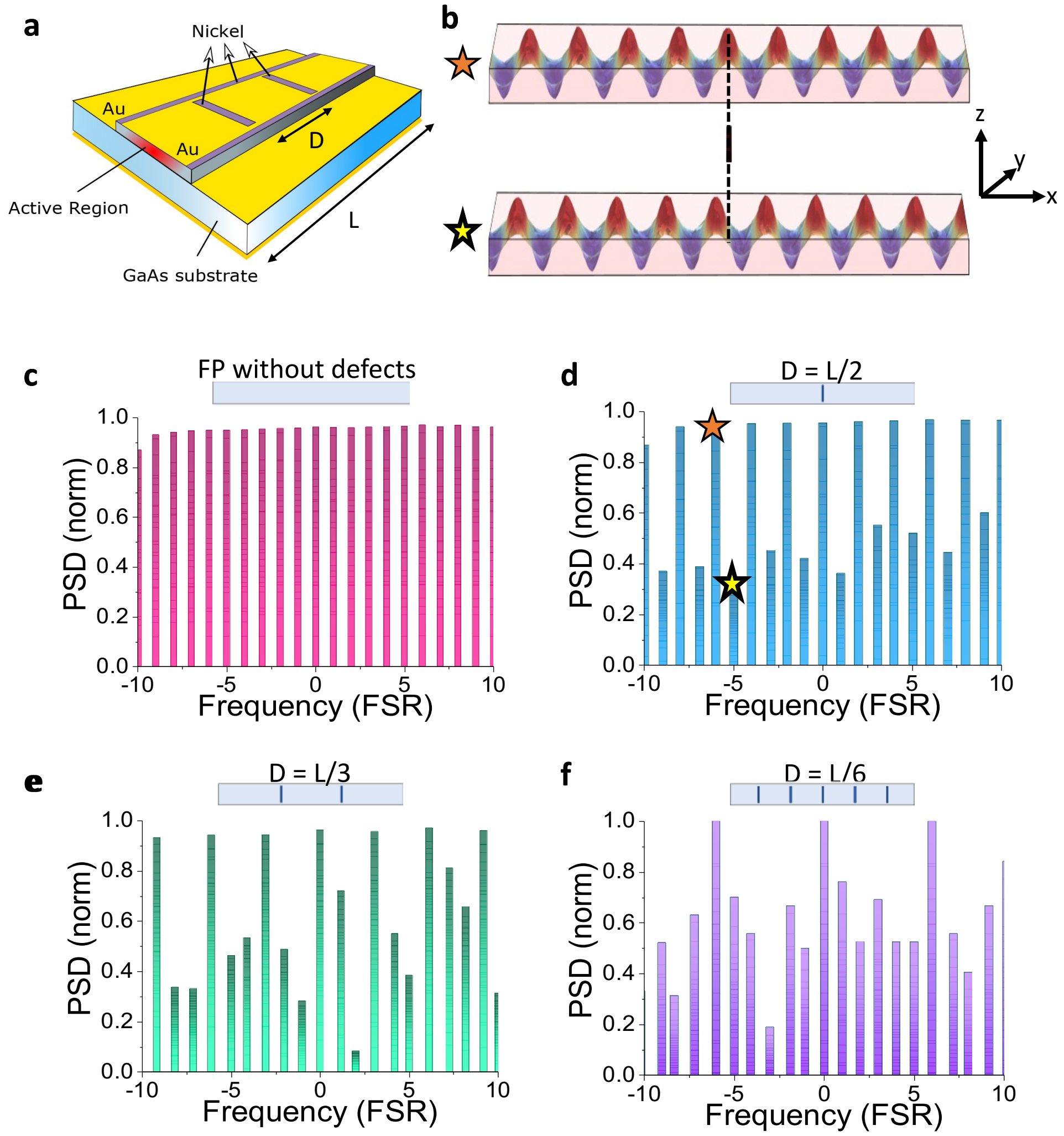

**Figure 1. Device concepts and simulated frequency dependence of the power spectral density. a** Three-dimensional device structure. The top contact of a double metal QCL waveguide is patterned with a periodic number of slits covered by a 5 nm Nickel layer. To obtain operation on the $n$-harmonic state, the distance between the slits is set to $L/(n+1)$. **b** Schematic diagram of the electric field distribution inside a QCL cavity with one defect placed at L/2. The upper and lower panels represent two neighbouring modes of the comb. The black rectangle and dashed line indicates the position of the defect on the top contact of the cavity. Changing the frequency also changes the relative position of the field peaks with respect to the defect. **c-f** Simulated power spectral density (PSD) of the field for different cases: a FP cavity without defects (**c**) shows a PSD equally distributed between all the modes; in the presence of defects (**d-f**), the frequencies for which the field peaks are not aligned with the defects show more significant losses resulting in a spectrum with a harmonic modulation. All spectra are centred with respect to 3 THz, and the frequency axis is normalized by the cavity FSR. The insets show the number and distribution of defects on the top contact of the cavity. The orange and yellow stars indicate, respectively, the case in which the defect is aligned with the field peaks and the one where it is misaligned (**b**).

A set of QCLs were then fabricated with the number of defects $n$ varying in the range 0–5. We employed a 17-μm-thick GaAs/AlGaAs heterostructure with an active region comprising three active modules, each tuned to a different central frequency, and each based on longitudinal optical phonon-assisted interminiband transitions. All stacks shows comparable threshold current densities ($J_{th}$),[32] and gain bandwidths centred at 2.5 THz, 3 THz and 3.5 THz (see Methods). The choice of a heterogeneous heterostructure guarantees the design of a harmonic spectral response broader than those achieved so far using only bias control in a single frequency laser.[27] The double-metal FP laser cavity includes top Ni-based lossy side absorbers to inhibit high-order lateral modes by increasing their threshold gain[34] (Methods). The defects were lithographically defined onto the top Au layer (Fig. 2a), exposing the active region below and then covered with a 5-nm-thick Ni layer (Fig. 2b). For consistency, all QCLs are fabricated with equal width 70 μm and length $L = 2.5$ mm.

The current-voltage (I-V) and light-current (L-I) characteristics of a prototypical QCL with $n = 5$ is reported in Figure 2c. The device has a threshold current density $J_{th}$=165 A/cm$^2$, 15% higher than that of the reference structure,[10] a consistent dynamic range $J_{max}/J_{th} = 3.5$, a maximum optical power of 3 mW, a slope efficiency of 16.5 mW/A, and a 0.05% wall-plug (WP) efficiency. A comparison of the main figures of merit of the complete set of fabricated devices is provided in the Supporting Information. Compared to the reference FC QCL, the HFCs show a 4–20% increase in the threshold current density ($J_{th}$), a corresponding 5–35% increase of the slope efficiency, a 20–29% variation of the WP efficiency, and a 16–22% variation of the peak optical power (Table II in Supporting Information). This proves that the defects redistribute the power of the suppressed modes into the harmonic modes, partially impacting the overall device performance. However, since there is no clear correlation with the number of defects implemented, such performance variations between the different HFCs are mostly device dependent.

The Fourier transform infrared (FTIR) emission spectra (Bruker, Vertex 80) taken under vacuum of the reference device (Fig. 2d), without defects, shows a FC regime with a maximum spectral coverage of 0.79 THz with a set of 50 optical modes equally spaced by 1 FSR = 15.7 GHz. When defects are implemented on the top of the laser cavity, the FTIR spectra show (Figs 2e-g) a pure harmonic regime, with a repetition rate that precisely follows the rationale of our simulation design, e.g. $n$ defects correspond to a HFC with a mode spacing equal to ($n$+1) FSR, as shown by the FTIR spectra and by the corresponding intermode beatnotes (Figs. 2i-j).

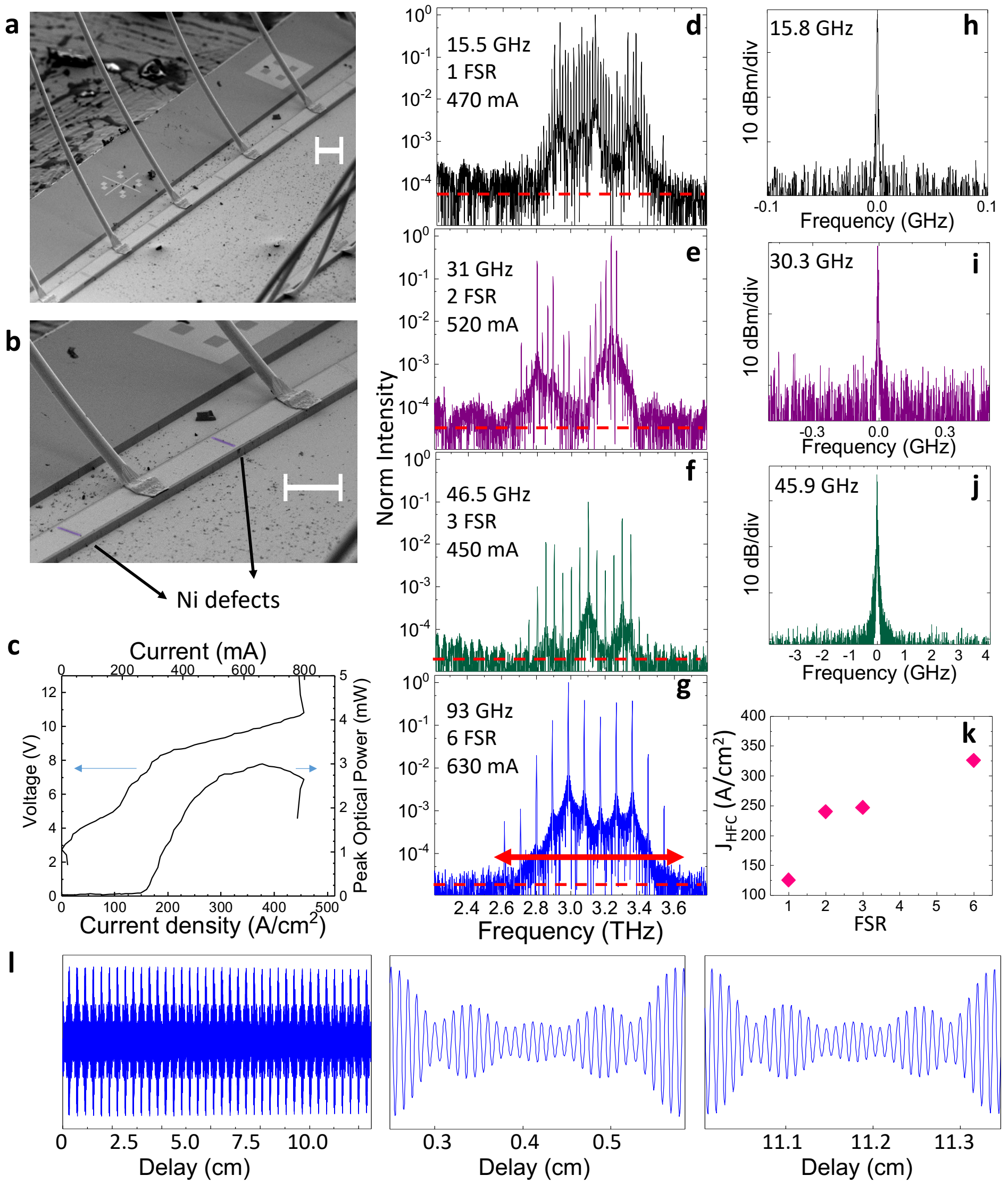

**Figure 2 Harmonic frequency comb spectra and intermode beatnote linewidth a-b.** Scanning electron microscope (SEM) image of a double metal QCL with the defects patterned on the top contact. The scale bars are 100 µm **c.** Voltage–Current density (V–J) and Light–Current density (L–J) characteristics of the HFC with $n$= 5, measured at 15 K. The peak optical power was measured using a broad-area THz absolute power meter (TK Instruments, aperture 55 × 40 mm$^2$). **d** FTIR emission spectrum of a non-patterned reference QCL acquired under vacuum in continuous wave, at 20 K with a 0.075 cm$^{-1}$ spectral resolution **e-g** FTIR spectra of three devices with the same dimensions as the reference laser of panel **d**, but with an increasing number of defects on the top contact ($n$= 1(e), $n$= 2(f), $n$= 5(g)). The dashed lines mark the noise level. The spectra become perfectly harmonic, according to the distance between the defects (2, 3 and 6 FSR). All spectra are collected at 20 K. **h-j** Intermode beatnote of the reference laser **(h)** and of the device with $n$ = 1 (e) and $n$ = 2 (f), recorded with an RF spectrum analyzer (Rohde and Schwarz FSW43, RBW: 300 Hz, video bandwidth (VBW): 300 Hz, sweep time (SWT): 20 ms, RMS acquisition mode), under the same conditions of the spectra d-f. **k.** Threshold current density of the HFC regime as a function of the harmonic order (FSR). **l** Interferogram (IFG) of the harmonic comb with $n$ = 5 corresponding to the spectrum (g), with two magnified views, one close to zero and one close to the maximum delay of the FTIR, showing the preservation of its symmetry.

In a free-running THz QCL, the FC regime is self-starting, and persists only in specific operational regimes, usually just above laser threshold,[9,33,35] in which the group velocity dispersion[36–38] is low enough. The gain medium however entangles the dispersion dynamics at higher driving currents resulting in FC operation usually occurring over less than 25 % of the lasing bias range.[32] Without a strategy to handle such a bias-dependent dispersion compensation, that includes active and passive approaches[39–43] or strategies to modulate the intra-cavity field,[10] the achievable spectral range of the FC is limited with respect to that set by the gain medium design. This is the case of the reference structure of the present work (Fig. 2d), in agreement with previous reports.[10,33] However, the devised HFCs show a different behavior: the current density threshold of the HFC regime $J_{HFC}$ increases as a function of $n$ as shown in Fig. 2k. This is in agreement with previous experimental reports in self-starting homogeneous THz HFCs[27] and consistent with the increased losses found at increasing $n$ (see Supporting Information, Fig. S3c). Below $J_{HFC}$, the spectra appear mostly single mode.

Interestingly, and in contrast to the reference laser, a stable HFC regime is found progressively more towards the driving current region in which the gain is clamped (530–750 mA)[44]. Here, the laser delivers its maximum optical power and the spectral coverage becomes hence broader, turning from 0.8 THz ($n$ = 1, Fig. 2e) to 0.85 THz ($n$ = 2, Fig. 2f) and then to 1.1 THz ($n$ = 5, Fig. 2g), demonstrating that the QCL operates as a stable self-starting FC without any of the aforementioned dispersion compensation strategies needing to be implemented. Remarkably the device (Fig. 2e, 2g), deliver a maximum optical power per mode of ~270 μW, the maximum value achieved so far in any THz QCL FC. This is reflected in the plots of the intermode beatnote (BN) as a function of driving current (Figs. 3a-c) that we detected via a free space antenna using a radio frequency spectrum analyzer equipped with a 90 GHz mixer (Rohde and Schwarz FS-Z90) to extend its operational range up to 90 GHz. In the case of the reference laser (Fig. 2d), a single, sharp (~40 dBm) and narrow linewidth (LW) ~ 0.7-10 kHz) BN, at ~15.7 GHz, persists for a current range of 100 mA (i.e. from 385 to 490 mA) (13% operational range), in agreement with previous reports (Fig. 3a).[33] For the HFC with $n$=1, the ~30.3 GHz BN (2FSR), having a signal-to-noise ratio of 28 dBm, persists for ~100 mA (i.e. from 420 to 520 mA) (Fig. 3b) and shows a narrow LW (1–10 kHz). The HFC with $n$=2 shows a stable and narrow (10-20 kHz) BN (~35 dBm) at ~45.7 GHz (3FSR) persisting over ~80 mA (i.e. from 430 to 500 mA) (Fig. 3c). It is worth mentioning that, in both cases, and in the investigated current range, no other beating across the RF spectrum is observed except the one of the harmonic beatnote, indicating the purity of the harmonic state. This was proved by inspecting the RF spectra at the subharmonic frequencies, with a lower resolution bandwidth (RBW) and a lower frequency span.

For the HFC with $n$=5 we could not collect the corresponding BN due to limitation of our spectrum analyzer. However, the coherence of the harmonic state can be verified to a certain degree by comparing the symmetry and periodicity of the measured interferogram envelope. Indeed, it has been shown[27] that preservation of symmetry in the interferogram. in each period throughout the travel range of the FTIR, is a necessary condition and an accessible tool to assess the presence of a pure comb state. Figures 2l show the interferogram of the HFC with $n$=5, corresponding to the spectrum 2g. The two magnified plots, around the zero-path delay and close to the maximum travel range, respectively, show the required symmetry preservation.

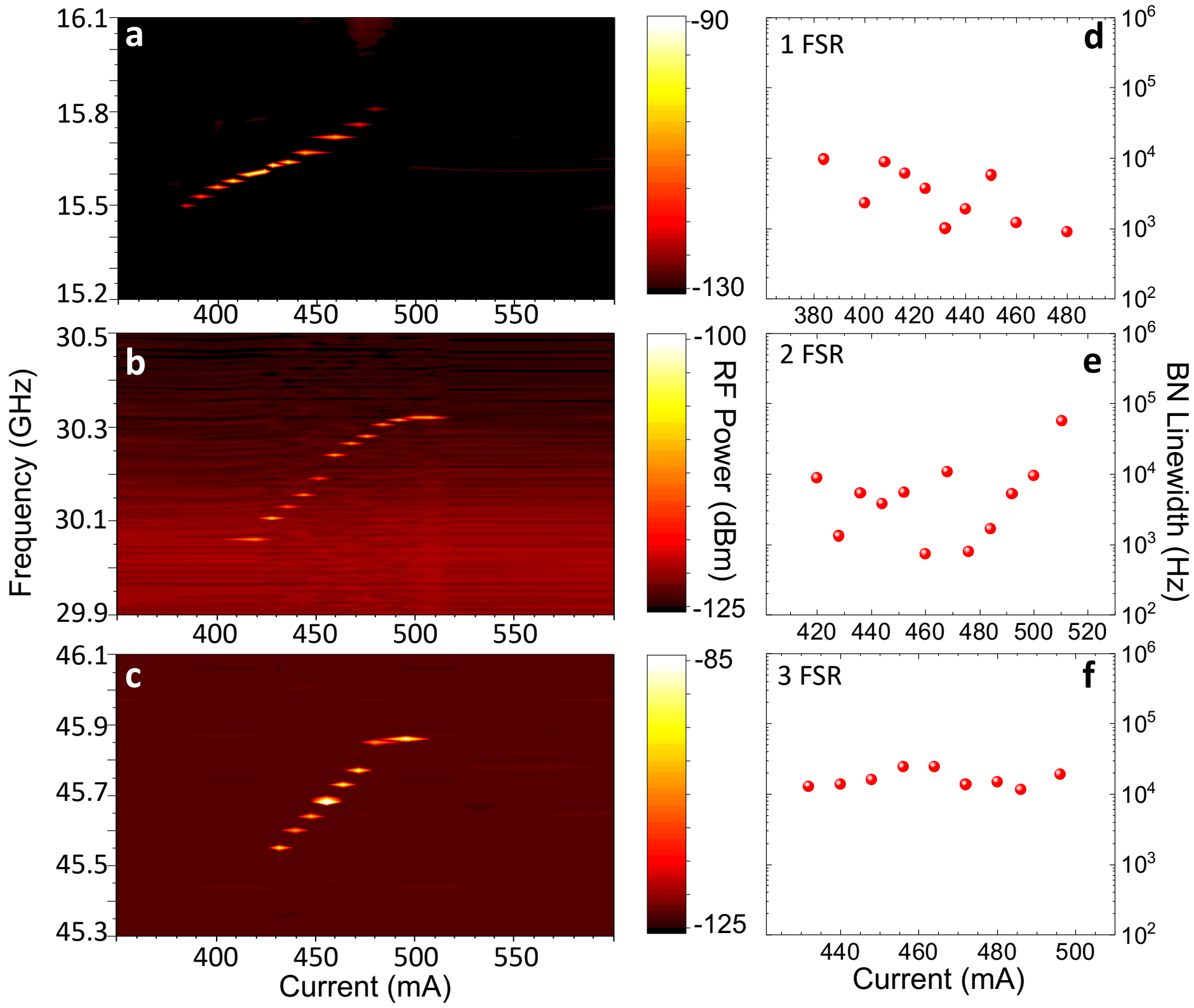

**Figure 3 Intermode beatnote maps and linewidth. a-c** Intermode beatnote maps as a function of the driving current, measured at 20 K for the QCL FC designed without defects **(a)**, with $n$ = 1 **(b)** and $n$ = 2 **(c)**. **d-f** BN linewidths as a function of the driving current, measured at 20 K for the QCL FC designed without defects **(d)**, with $n$ = 1 **(e)** and $n$ = 2 **(f)**.

The same device also exhibits a harmonic-like regime at 3 FSR that develops from 400 to 500 mA (see Supplementary information). This behavior is expected, considering the proposed design. In fact, the $n = 5$ defects include the $n = 2$ defects of the 3 FSR harmonic design. These two defects induce, in a separate current regime, the device to also develop this submultiple harmonic-like behavior, before reaching the threshold of the sixth harmonic comb regime. In this regime, the spectrum appears nearly harmonic and the corresponding IFG does not preserve its symmetry throughout the travel range of the FTIR (see Supplementary information).

We next investigate the dynamical processes underlying harmonic comb generation by performing time-domain simulations, employing a Maxwell-density matrix approach.[45–47] The optical propagation equations for the waveguide field components are here coupled to a density matrix description of the active region dynamics (see Methods). The waveguide parameters used can be found in Table I.

| Parameter | Value |
|---|---|
| Loss coefficient $a_0$ | $(12.75 + 0.25 \times n)$ cm$^{-1}$ |
| Overlap factor | 1 |
| Center frequency | 3.0 THz |
| Mirror reflectivity (power) | 70.88 % |
| Slit reflectivity (power) | 0.04 % |
| Refractive index | 3.66 |
| GVD coefficient $\beta_2$ | $1.0\times10^{-22}$ s$^2$m$^{-1}$ |

**Table I. Simulation parameters.** Waveguide parameters for the Maxwell-density matrix simulations. $n$ denotes the number of defects.

In Fig.4a-e, the simulated power spectra for $n = 1,2,3,5$ and 7 equally spaced slits in the waveguide are shown, along with the corresponding RF beatnote. As seen experimentally, the harmonic order of the comb directly scales with the number of slits, i.e., the comb line spacing and thus the RF beatnote frequency is related to the round-trip time $t_{rt}$ via $\Delta f = (n+1)/t_{rt}$. In all cases, narrow beatnotes with widths below the numerical resolution limit are obtained, indicating phase-locked comb operation in agreement with the experimental results.

In Fig. 4f-j, the corresponding time traces of the outcoupled power are shown, featuring a significant amplitude modulation at the corresponding harmonic of the round-trip time. To obtain the harmonic combs of different orders shown in Fig. 4, we had to adapt the currents by slightly changing the doping.

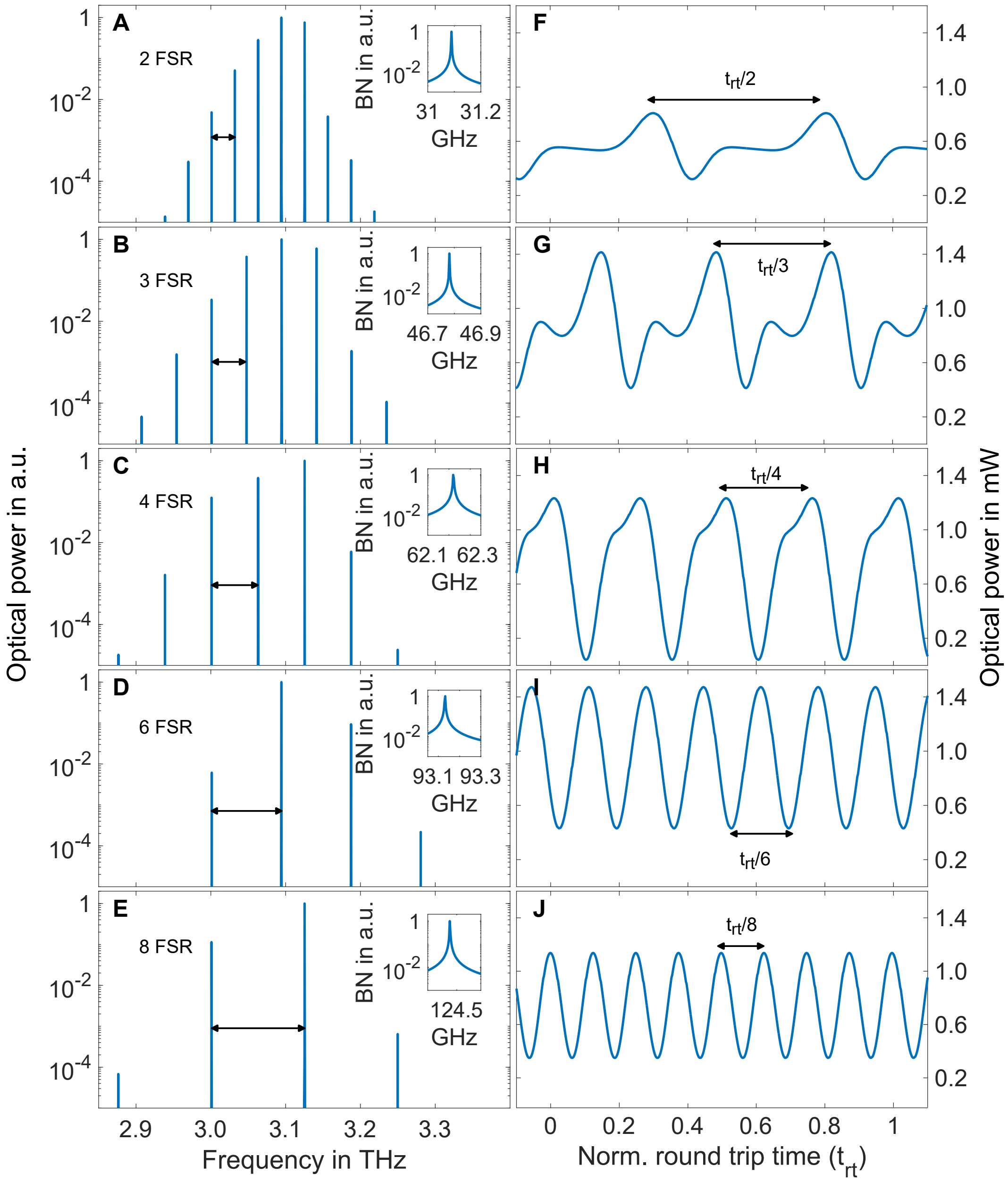

**Figure 4: Dynamical simulation results using the Maxwell-density matrix formalism. a-e,** Power spectra calculated from the last 5000 round trips, for $n$ = 1,2,3,5,7 slits. The spacing between the individual comb lines closely follows the number of defects. Insets: RF beatnote (BN). The BN frequency increases proportionally with the order of the harmonic comb. The BN width is limited by the numerical frequency resolution (~1.56 MHz), indicating phase-locked comb operation for each device. **f-j,** Time traces of the outcoupled optical power for the corresponding spectra shown in the left column. Significant amplitude modulations are visible, with a periodicity corresponding to $t_{rt}$ divided by the order of the harmonic comb.

For example, Fig. 4a ($n$= 1) was obtained for 466 mA, and Fig. 4e ($n$= 7) for 487 mA. This coincides with experimental observations that the current required for harmonic comb operation gradually increases with the number of slits. Also the spectral coverage becomes broader, in agreement with above experimental results. For currents below the harmonic comb regime, our simulations yield single mode lasing, and for currents above the harmonic comb regime, unlocked multi-mode operation with a broad beatnote is obtained, in agreement with the beatnote maps of Fig. 3. These results indicate that the defects introduce an additional resonance condition in the waveguide, favouring the respective modes and driving the field into a periodic waveform at the corresponding harmonic of the fundamental repetition rate.

For too high intensities, however, other mechanisms such as the Risken-Nummedal-Graham-Haken instability start to overcome the effect introduced by the slits, destroying the phase coherence between the modes[48].

By implementing a defined number of defects on the top contact of a wire THz laser, we introduce a spatially periodic modulation of the optical gain that promotes the emission of HFC states. The purely harmonic regimes, developing at bias currents close to the peak of the laser emission, shows a spectral bandwidth significantly larger (1.1 THz) than that found on the free running fundamental harmonic (1FSR) THz QCL combs. The ability of the laser to distribute its optical power among a few, powerful modes separated by many FSRs (as opposed to many, weaker adjacent cavity modes) opens up previously unforeseen possibilities for QCLs for arbitrary optical and microwave waveform generation[49] or for mode locking[50]. Indeed, provided that the repetition rate is on the order of the inverse of the gain recovery time or faster, the QCL can no longer be considered a fast saturable gain medium in the HFC regime, as was previously assumed.[26] This is also extremely appealing for broadband spectroscopy. The advantage of HFC over FC is that the higher power per mode and broader spacing characteristic of this state are expected to lead to a greater mode proliferation, and thus ultimately to a broader spectral coverage.[50] Furthermore, HFCs open intriguing perspectives for THz quantum key distribution of interest in quantum technologies or for the generation of terahertz signals of high spectral purity in wireless communication networks.[13]

## Methods

### Active region design

The individual active regions consist of nine GaAs quantum wells, forming a cascade of alternating photon and LO-phonon-assisted transitions between two quasi-minibands[51]. Each active region module shows a diagonal laser transition. The layer sequences of the three active regions designs are: 10.6/**0.5**/17/**1**/13.5/**2.1**/12.4/**3.1**/10/**3.1**/9/**2.9**/7.5/**3.1**/17.8/**3.1**/15.2/**4.1**;9.8/**0.5**/15.7/**1**/12.5/**2**/11.4/**2.9**/9.2/**2.9**/8.3/**2.9**/6.9/**2.9**/16.5/**2.9**/14.1/**3.9**; and 10.5/**0.5**/12.2/**1**/12.5/**1.9**/11/**2.8**/8.8/**2.8**/7.9/**2.8**/6.6/**2.8**/15.8/**2.8**/13.8/**3.7**, where the thickness is in nanometres, the bold figures represent the AlGaAs barriers, and the underlined figures denote the Si-doped ($3.5 \times 10^{16}$ cm$^{-3}$) GaAs layer. The aluminum content in the barriers is 0.14, 0.16, and 0.18, respectively.

## QCL fabrication

Fabry-Perot laser bars are fabricated in a metal–metal waveguide configuration via Au-Au thermo-compression wafer bonding of the 17-$\mu$m-thick active region (sample L1494) onto a highly doped GaAs substrate, followed by the removal, through a combination of mechanical lapping and wet etching, of the host GaAs substrate of the molecular beam epitaxial (MBE) material. The $Al_{0.5}Ga_{0.5}As$ etch stop layer is then removed using HF etching. Vertical sidewalls are defined by inductively coupled plasma etching of the laser bars to provide uniform current injection. A Cr/Au (10 $\mu$m/150 $\mu$m) top contact, patterned with 2×42 μm rectangular open slits, is then deposited along the center of the ridge surface, leaving a 3-$\mu$m-wide region uncovered along the ridge edges. A thin layer of Ni (5-nm-thick) was then deposited over the defects and over the uncovered region (side absorbers) using a combination of optical lithography and thermal evaporation. This lossy Ni layer on the side absorbers is intended to inhibit lasing of the higher order lateral modes by increasing their threshold gain.[52] Finally, the backside of the substrate is lapped down to 150 $\mu$m to allow a proper thermal management and enable operation in CW. Laser bars, 70 $\mu$m wide and 2.5 mm long, are then cleaved and mounted on a copper bar, wire bonded, and then mounted onto the cold finger of a He continuous-flow cryostat.

## Model

The harmonic comb operation is modeled using a coupled Maxwell-density matrix approach,[45] which is combined with carrier transport simulations to avoid free fitting parameters.[46] This approach has been found highly suitable for quantitative modeling of mode-locked QCL operation.[46,47] Here, the quantum active region dynamics is captured by a Lindblad-type density matrix equation for a representative QCL period, supplemented by periodic boundary conditions.[45,46] Similarly as for previous simulations of a short-pulse structure based on the same active region design, the eigenenergies, optical dipole moment of the laser transition and scattering/dephasing rates are obtained from carrier transport simulations of the 2.5 THz stack[47]. The resulting model system comprises 9 energy levels. Employing the rotating wave/slowly varying amplitude approximation, the optical field propagation is described by[45]

$$v_g^{-1}\,\partial_t E^{\pm} \pm \partial_z E^{\pm} = p^{\pm} - aE^{\pm} - \mathrm{i}\frac{\beta_2}{2}\partial_t^2 E^{\pm}, \qquad (1)$$

with the complex electric field amplitudes $E^{\pm}(x,t)$ of the forward/backward traveling component. Here, $z$ and $t$ denote the position in the waveguide and time. Furthermore, $v_g$ is the group velocity, $p^{\pm}(z,t)$ represents the polarization term computed from the density matrix,[45,46] $a$ is the power loss coefficient, and $\beta_2$ denotes the background group velocity dispersion (GVD) coefficient. The waveguide slits are modeled using reflection and transmission coefficients $r$ and $t$, with

$$E_L^- = rE_L^+ + tE_R^-, E_R^+ = rE_R^- + tE_L^+. \qquad (2)$$

Here, the subscripts L and R refer to the field components directly to the left and right of the slit. Each simulation runs over more than 10,000 roundtrips to ensure convergence to steady state.

**Acknowledgements** This work was supported by the European Union through the HORIZON-CL4-2021-DIGITAL-EMERGING-01 (101070546) Muquabis, the QuantERA 2021 project QATACOMB (Project No. 491 801 597), by the Italian Ministry of University and Research through the PNRR project PE0000023-NQSTI, and by the EPSRC (UK) programme grant 'TeraCom' (EP/W028921/1).

**Author contributions**: M.S.V. and E.R. conceived the concept. E.R. fabricated the devices, set up the transport and optical experiment. E.R., M.A.J.G and V.P performed numerical simulations and interpreted the data. L.L, A.G.D. and E.H.L grew by molecular beam epitaxy the QCL structure. C.J. and L. L. developed the theoretical model. The manuscript was written by M.S.V. and E.R. M.S.V. coordinated and supervised the project. All authors contributed to the final version of the manuscript and to discuss the results.

**Supplementary Information Available**

The following files are available free of charge. SupportingInfo.pdf

**Competing financial interests:**

The authors declare no competing financial interests.

**Data availability:** The data presented in this study are available on reasonable request from the corresponding author.

**Code availability statement:** The relevant computer codes supporting this study are available from the authors upon reasonable request.